\documentclass[showpacs,prc,preprint,nofootinbib,showkeys]{revtex4}
\pdfoutput=1

\usepackage{graphicx}
\usepackage{dcolumn}
\usepackage{bm}
\usepackage{slashed}
\usepackage{amsmath,graphicx}
\usepackage[colorlinks=true,linktocpage=true,linkcolor=blue,citecolor=blue]{hyperref}
\usepackage{float}
\usepackage{nicefrac}
\usepackage[normalem]{ulem}
\usepackage{amsmath}
\usepackage{subfigure}

\usepackage{bbold}

\newcommand{\beq}{\begin{eqnarray}}
\newcommand{\eeq}{\end{eqnarray}}

\newcommand{\bel}[1]{\begin{eqnarray}\label{#1}}
\newcommand{\eel}{\end{eqnarray}}

\newcommand{\rf}[1]{Eq.~(\ref{#1})}

\newcommand{\rftwo}[2]{Eqs.~(\ref{#1})~and~(\ref{#2})}
\newcommand{\rfn}[1]{~(\ref{#1})}

\newcommand{\nn}{\nonumber}

\newcommand{\trt}{{\rm tr_2}}
\newcommand{\trf}{{\rm tr_4}}

\newcommand{\f}[2]{\frac{#1}{#2}}
\newcommand{\onehalf}{{\nicefrac{1}{2}}}

\renewcommand\sout{\bgroup \color{blue} \ULdepth=-.5ex \ULset}

\def\TmnU{T^{\mu \nu}}

\def\n0{n_{(0)}}
\def\e0{\varepsilon_{(0)}}
\def\P0{P_{(0)}}
\def\s0{s_{(0)}}

\def\fplusrsxp{f^+_{rs}(x,p)}

\def\fminusrsxp{f^-_{rs}(x,p)}

\def\bmu{\beta_\mu}

\def\umU{u^\mu}  
                         
\def\umL{u_\mu}

\def\unu{u^\nu}

\def\unuL{u_\nu}

\def\omnL{\omega_{\mu\nu}}
\def\omnU{\omega^{\mu\nu}}
\def\omnLbar{{\bar \omega}_{\mu\nu}}
\def\omnUbar{{\bar \omega}^{\mu\nu}}

\def\oabL{\omega_{\alpha\beta}}
\def\oabU{\omega^{\alpha\beta}}
\def\omnLD{{\tilde \omega}_{\mu\nu}}
\def\omnUD{\tilde {\omega}^{\mu\nu}}
\def\omnLDbar{{\bar {\tilde \omega}}_{\mu\nu}}
\def\omnUDbar{{\bar {\tilde {\omega}}}^{\mu\nu}}

\def\epsLmnab{\epsilon_{\mu\nu\alpha\beta}}

\def\epsLmnab{\epsilon_{\mu\nu\alpha\beta}}

\def\epsLijk{\epsilon_{ijk}}

\def\pmu{p^\mu}
\def\pnu{p^\nu}

\def\vv{{\boldsymbol v}}
\def\pv{{\boldsymbol p}}

\def\piv{{\boldsymbol \pi}}

\def\Pv{{\boldsymbol P}}
\def\Pbarv{{\bar {\boldsymbol P}}}
\def\calP{{\cal P}}
\def\calPv{{\boldsymbol {\cal P}}}

\def\uv{{\boldsymbol u}}

\def\kv{{\boldsymbol k}}

\def\kmL{k_\mu}
\def\kbar{\bar k}
\def\kmLbar{{\bar k}_\mu}

\def\knL{k_\nu}

\def\ov{{\boldsymbol \omega}}
\def\obar{{\bar \omega}}

\def\obar{\bar\omega}

\def\omL{\omega_\mu}

\def\omLbar{{\bar \omega}_\mu}

\def\omb{\omega^\beta}

\def\ev{{\boldsymbol e}}
\def\bv{{\boldsymbol b}}

\def\gmU{\gamma^\mu}

\def\SmunuU{{\Sigma}^{\mu\nu}}

\def\SabU{{\Sigma}^{\alpha\beta}} 
\def\S0iU{{\Sigma}^{0i}} 
\def\SijU{{\Sigma}^{ij}} 
\def\SmnU{{\Sigma}^{\mu\nu}}

\def\ubarrp{{\bar u}_r(p)}

\def\usp{u_s(p)}
\def\urp{u_r(p)}

\def\vbarrp{{\bar v}_r(p)}
\def\vbarsp{{\bar v}_s(p)}
\def\vsp{v_s(p)}
\def\vrp{v_r(p)}

\def\g5{\gamma_5}

\def\slmnU{S^{\lambda, \mu \nu}}

\def\Ot{\tilde \Omega}

\def\sigUk{\sigma_{rs}^k}
\def\sigv{{\boldsymbol \sigma}}
\def\sigvrs{{\boldsymbol \sigma}_{rs}}
\def\sv{{\boldsymbol \sigma}}

\newcommand{\lp}{\left(}
\newcommand{\rp}{\right)}

\newcommand{\lsb}{\left[}
\newcommand{\rsb}{\right]}

\newcommand{\zv}{{\boldsymbol z} }

\newcommand{\SL}{{\cal S}}

\begin{document}
 

\title{Spin-dependent distribution functions for relativistic\\ hydrodynamics of spin-$\onehalf$ particles}

\author{Wojciech Florkowski} 
\affiliation{Institute of Nuclear Physics Polish Academy of Sciences, PL-31342 Krakow, Poland}
\affiliation{Jan Kochanowski University, PL-25406 Kielce, Poland}
\affiliation{ExtreMe Matter Institute EMMI, GSI, D-64291 Darmstadt, Germany}
\author{Bengt Friman} 
\affiliation{GSI Helmholtzzentrum f\"ur Schwerionenforschung, D-64291 Darmstadt, Germany}
\author{Amaresh Jaiswal} 
\affiliation{School of Physical Sciences, National Institute of Science Education and Research, HBNI, Jatni-752050, India}
\affiliation{ExtreMe Matter Institute EMMI, GSI, D-64291 Darmstadt, Germany}
\author{Radoslaw Ryblewski} 
\affiliation{Institute of Nuclear Physics Polish Academy of Sciences, PL-31342 Krakow, Poland}
\affiliation{ExtreMe Matter Institute EMMI, GSI, D-64291 Darmstadt, Germany}
\author{Enrico Speranza} 
\affiliation{Institute for Theoretical Physics, Goethe University, D-60438 Frankfurt am Main, Germany}
\affiliation{GSI Helmholtzzentrum f\"ur Schwerionenforschung, D-64291 Darmstadt, Germany}
\date{\today}

\date{\today}


\bigskip

\begin{abstract}
Recently advocated expressions for the phase-space dependent spin-$\onehalf$ density matrices of particles and antiparticles are analyzed in detail and reduced to the forms linear in the Dirac spin operator. This allows for a natural determination of the spin polarization vectors of particles and antiparticles by the trace of products of the spin density matrices and the Pauli matrices. We demonstrate that the total spin polarization vector obtained in this way agrees with the Pauli-Luba\'nski four-vector, constructed from an appropriately chosen spin tensor and boosted to the particle rest frame. We further show that several forms of the spin tensor used in the literature give the same Pauli-Luba\'nski four-vector.
\end{abstract}



\maketitle 

\section{Introduction}
\label{sec:i}

The idea that the global angular momentum of the hot and dense matter created in heavy-ion collisions may be reflected in the polarization of  $\Lambda$ hyperons and vector mesons, has triggered broad interest in studies of the possible relation between vorticity and polarization~\cite{Liang:2004ph, Liang:2004xn, Betz:2007kg, Becattini:2007sr, Becattini:2013vja, Becattini:2016gvu, Pang:2016igs, Abelev:2007zk, STAR:2017ckg} (see Ref.~\cite{Wang:2017jpl} for a recent review). The study of vorticity has gained widespread interest also because it is an important ingredient in studies of theories that deal with the production of false QCD vacuum states and chiral symmetry restoration~\cite{Kharzeev:2015znc}. Theoretical studies of vorticity, polarization and related topics have explored the role of the spin-orbit coupling~\cite{Liang:2004ph, Betz:2007kg, Gao:2007bc, Chen:2008wh}, the polarization of rigidly rotating fluids in global equilibrium~\cite{Becattini:2007sr, Becattini:2009wh, Becattini:2013fla}, the kinetics of spin~\cite{Gao:2012ix, Fang:2016vpj, Fang:2016uds}, and anomalous hydrodynamics~\cite{Son:2009tf, Kharzeev:2010gr}. We also note the recent work based on the Lagrangian formulation of hydrodynamics~\cite{Montenegro:2017rbu, Montenegro:2017lvf}. 

Indeed, in non-central heavy-ion collisions, a fireball is created with large global angular momentum, which may generate spin polarization in a way that resembles the Einstein-de Haas~\cite{dehaas:1915} and Barnett~\cite{Barnett:1935} effects. Since such collisions are well described by relativistic hydrodynamic models~\cite{Gale:2013da, Jaiswal:2016hex, Florkowski:2017olj}, it is of interest to include polarization explicitly in hydrodynamics. So far, polarization effects have been taken into account only at the end of the hydrodynamic expansion, i.e., on the freeze-out hypersurface where a connection between vorticity and polarization was assumed~\cite{Becattini:2009wh, Becattini:2013fla}. In such approaches, the preceding dynamics of the polarization, from the initial stages of the collision until the freeze-out, is not accounted for.

Recently, a new hydrodynamic framework was constructed~\cite{Florkowski:2017ruc}, which fully incorporates spin-degrees of freedom in a perfect-fluid approach. This approach is based on the local-equilibrium, spin dependent phase-space distribution functions $f^\pm(x,p)$, put forward in Ref.~\cite{Becattini:2013fla}. In this work, we study formal aspects connected with the calculation of thermodynamic and hydrodynamic quantities using the functions $f^\pm(x,p)$. We reduce the original exponential form to an expression linear in the Dirac spin operator $\SmunuU = \f{i}{4} [\gamma^\mu,\gamma^\nu]$. This allows for a straightforward determination of the spin-polarization vectors of particles and antiparticles by evaluating the trace of the product of the phase-space densities with Pauli matrices. We show that the total spin-polarization vector obtained in this way agrees with the Pauli-Luba\'nski (PL) four-vector~\cite{Lubanski:1942}~\footnote{In the particle rest frame, the PL four-vector does not change sign under reflections and is, therefore, often called the pseudo four-vector.}, constructed from the spin tensor used in~\cite{Florkowski:2017ruc} and boosted to the particle rest frame. Interestingly, other forms of the spin tensors used in the literature yield the same PL four-vector (except for the Belinfante construction which sets the spin tensor equal to zero). This indicates that the form used in \cite{Florkowski:2017ruc} represents an appropriate classical approximation for the spin tensor. 

{\it Conventions and notation:} We use the following conventions and notation for the metric tensor, the four-dimensional Levi-Civita's tensor, and the scalar product in Minkowski space: $g_{\mu\nu} = \hbox{diag}(+1,-1,-1,-1)$, $\epsilon^{0123} = -\epsilon_{0123} = 1$, $a^\mu b_\mu= g_{\mu \nu} a^\mu b^\nu$. Three-vectors are shown in bold font and a dot is used to denote the scalar product of both four- and three-vectors, e.g., $a^\mu b_\mu = a \cdot b = a^0 b^0 - {\boldsymbol a} \cdot {\boldsymbol b}$. For the three-dimensional Levi-Civita tensor $\epsilon_{ijk}$, with $\epsilon_{123}=+1$, we do not distinguish between lower and upper components, note that $\epsilon_{0123} = -\epsilon_{123} = -1$. The symbol $\mathbb{1}$ is used for a two-by-two or four-by-four unit matrix. On the other hand, we distinguish the trace in the spin and spinor spaces by using the symbols $\trt$ and $\trf$, respectively.

The components of the four-momentum of a particle with the mass $m$ are $p^\mu = (E_p, \pv)$, with $E_p$ being the  on-mass-shell energy, $E_p = \sqrt{m^2 + \pv^2}$, and the components  of the four-velocity of the fluid element are  $u^\mu = (u^0, \uv)$. The quantities defined in the particle rest frame (PRF) are marked by an asterisk, those defined in the local fluid rest frame (LFRF) are labeled with a circle, while unlabeled quantities refer to the laboratory frame (LAB). Using this convention, the symbol $\uv_\ast$ denotes the components of the fluid three-velocity seen in the particle rest frame, whereas $\pv_\circ$ denotes the components of a particle three-momentum in the local fluid rest frame.~\footnote{For a particle with four-momentum $p$ in the laboratory frame, the particle rest frame is boosted from LAB by the three-velocity $\vv_p =\pv/E_p$, while the local fluid rest frame is boosted from LAB by  $\vv = \uv/u^0$. The boosts considered in this work are all canonical or pure boosts~\cite{Leader:2001}. Their explicit form is given in Sec.~\ref{sec:pboost}} 

The sign and label conventions for the Dirac bispinors are given in Appendix~\ref{sec:DS}. Except for Appendix~\ref{sec:spinden}, where we temporarily switch to the chiral representation, all calculations are done using the Dirac representation for the gamma matrices. Throughout the text we use natural units with $c = \hbar = k_B =1$.

\section{Spin dependent distribution functions}
\label{sec:spinden}

\subsection{Basic definitions}
\label{sec:basdef}

In this work we analyse the phase-space distribution functions for spin-$\onehalf$ particles and antiparticles in local equilibrium, introduced in Ref.~\cite{Becattini:2013fla}. To include spin degrees of freedom, the standard scalar functions are generalized to two by two matrices in spin space for each value of the space-time position $x$ and four-momentum $p$,
\beq
\left[ f^+(x,p) \right]_{rs}  \equiv  \fplusrsxp &=&  \ubarrp X^+ \usp, \label{fplusrsxp}  \\
\left[ f^-(x,p) \right]_{rs}  \equiv \fminusrsxp &=& - \vbarsp X^- \vrp. \label{fminusrsxp}
\eeq
Here $\urp$ and $\vrp$ are Dirac bispinors (with the spin indices $r$ and $s$ running from 1~to~2), and the normalization $\ubarrp \usp=\,\delta_{rs}$ and $\vbarrp \vsp=-\,\delta_{rs}$. Note the minus sign and different ordering of spin indices in \rf{fminusrsxp} compared to \rf{fplusrsxp}~\footnote{To simplify the notation, the factors $2m$ appearing explicitly in the normalization conditions used in Refs.~\cite{Becattini:2013fla, Florkowski:2017ruc} (with $m$ being the particle mass), is here included in the definitions of bispinors.}. The objects $f^\pm(x,p)$ are two by two Hermitian matrices with the matrix elements defined by \rftwo{fplusrsxp}{fminusrsxp}.

Following Refs.~\cite{Becattini:2013fla, Florkowski:2017ruc}, we use the matrices
\bel{XpmM}
X^{\pm} =  \exp\left[\pm \xi(x) - \bmu(x) \pmu \right] M^\pm, 
\eel
where
\bel{Mpm}
M^\pm = \exp\left[ \pm \f{1}{2} \omnL(x)  \SmunuU \right] .
\eel
In \rftwo{XpmM}{Mpm}, $\beta^\mu= \umU/T$ and $\xi = \mu/T$, 
with the temperature $T$, chemical potential~$\mu$ and the fluid four velocity $\umU$ (normalised to unity). The quantity $\omnL$ is the polarization tensor. For the sake of simplicity, we restrict ourselves to classical Boltzmann statistics in this work.~\footnote{We note that by performing an analytic continuation of the polarization tensor, $\omnL\to -i \omnL$, the matrix $M^+$ becomes a representation of the Lorentz transformation $S(\Lambda)$ with $\Lambda^\mu_{\,\,\,\nu} = g^\mu_{\,\,\,\nu}+\omega^\mu_{\,\,\,\nu}$}

\subsection{Polarization tensor}
\label{sec:polten}

The antisymmetric polarization tensor $\omnL$ is defined by the tensor decomposition
\bel{omunuL}
\omnL \equiv \kmL \unuL - \knL \umL + \epsLmnab u^\alpha \omb,
\eel
where $k \cdot u = \omega \cdot u = 0$ and
\bel{kmuomu}
k_\mu = \omnL \unu, \quad \omega_\mu = \f{1}{2} \epsLmnab \, \omega^{\nu\alpha} u^\beta.
\eel
We note that  $k_\mu$ and $\omega_\mu$ are space-like four-vectors with only three independent components. In early works on fluids with spin \cite{Weyssenhoff:1947}, the so called Frenkel condition, $k_\mu = 0$, was introduced. We shall refer to this condition below.

The dual polarization tensor is defined by the expression
\bel{omunuLD}
\omnLD \equiv \f{1}{2} \epsLmnab  \oabU = \omega_\mu \unuL - \omega_\nu \umL +  \epsLmnab k^\alpha u^\beta.
\eel
Using \rftwo{omunuL}{omunuLD} one easily finds 
\bel{om2}
\f{1}{2} \omnL \omnU = k \cdot k - \omega \cdot \omega, \quad 
\f{1}{2} \omnLD \omnU = 2\, k \cdot \omega, \quad
\f{1}{2} \omnLD \omnUD =  \omega \cdot \omega - k \cdot k.
\eel

It is instructive to introduce another parameterization of the polarization tensor, which uses electric- and magnetic-like three-vectors in LAB, $\ev = (e^1,e^2,e^3)$ and $\bv = (b^1,b^2,b^3)$. In this case we write (following the sign conventions of Ref.~\cite{Jackson:1998})
\bel{omeb}
\omnL= 
\begin{bmatrix}
0       &  e^1 & e^2 & e^3 \\
-e^1  &  0    & -b^3 & b^2 \\
-e^2  &  b^3 & 0 & -b^1 \\
-e^3  & -b^2 & b^1 & 0
\end{bmatrix}.
\eel
Using \rf{omeb} in \rf{kmuomu} one finds
\beq
k^\mu &=& (k^0, \kv) =  \left( \ev \cdot \uv, u^0 \ev + \uv \times \bv \right), \nn \\
\omega^\mu &=& (\omega^0, \ov) =  \left( \bv \cdot \uv, u^0 \bv - \uv \times \ev \right) .
\label{komFROMeb}
\eeq
In the LFRF, where $u^0=1$ and $\uv=0$, we have $\kv=\ev$ and $\ov = \bv$ (i.e., $\kv_\circ=\ev_\circ$ and $\ov_\circ = \bv_\circ$). It is interesting to observe that $k^\mu$ plays the role of an electric field, while $\omega^\mu$ can be interpreted as a magnetic field acting on the magnetic moments. In order to switch from $\omnL$ to the dual tensor $\omnLD$, one replaces $\ev$ by $\bv$ and $\bv$ by~$-\ev$. Using \rf{komFROMeb}, one finds
\bel{om3}
\f{1}{2} \omnL \omnU =   \bv \cdot \bv - \ev \cdot \ev, \quad 
\f{1}{2} \omnLD \omnU = -2 \ev \cdot \bv , \quad
\f{1}{2} \omnLD \omnUD = \ev \cdot \ev - \bv \cdot \bv.
\eel

\subsection{Spin matrices $M^\pm$}
\label{sec:Mpm}

In Appendix~\ref{sec:spinden}, we show that the exponential dependence of the distribution function on $\Sigma^{\mu\nu}$ given in \rf{Mpm}, which is defined in terms of a power series, can be resummed. This results in an expression for $M^\pm$, linear in $\Sigma^{\mu\nu}$,
\beq
M^{\pm} &=&  \mathbb{1} \left[ \Re(\cosh z) \pm \Re\lp\frac{\sinh z}{2 z}\rp \omega_{\mu \nu}  \Sigma^{\mu \nu}\right]
\nonumber \\
&&\qquad\qquad+ i \gamma_5 \left[ \Im(\cosh z)\pm \Im\lp\frac{\sinh z}{2 z}\rp\omega_{\mu \nu}   \Sigma^{\mu \nu} \right],
\label{MpmExp}
\label{Mapp3} 
\eeq
where $\mathbb{1}$ is a unit matrix and
\beq 
z  &=&  \f{1}{2 \sqrt{2}} \sqrt{ \omnL \omnU + i \omnL \omnUD} 
= \f{1}{2} \sqrt{k \cdot k - \omega \cdot \omega + 2 i  k \cdot \omega} \,.
\label{z}
\eeq

It was demonstrated in Ref.~\cite{Florkowski:2017ruc} that a consistent thermodynamic description of particles with spin is obtained for real $z$. In this case $z$ can be interpreted as the spin chemical potential $\Omega$ divided by $T$. Here we follow this approach and restrict our considerations to the case where
\bel{conONE}
k \cdot \omega = \ev \cdot \bv = 0, \qquad k \cdot k - \omega \cdot \omega =  \bv \cdot \bv - \ev \cdot \ev \geq 0 \,.
\eel
Consequently, in what follows, we replace $z$ by a real number $\zeta$ in \rf{MpmExp} and use 
\bel{Mpmexp}
M^\pm &=& \cosh(\zeta) \pm  \f{\sinh(\zeta)}{2\zeta}  \, \omnL \SmunuU  ,
\eel
where
\bel{zeta}
\zeta = \f{\Omega}{T} =  \f{1}{2} \sqrt{ k \cdot k - \omega \cdot \omega } =
 \f{1}{2} \sqrt{ \bv \cdot \bv - \ev \cdot \ev } \,.
\eel
At this point it is convenient to introduce the rescaled quantities: 
\bel{rescaled}
\omnLbar  =  \f{\omnL}{2 \zeta}, \qquad
\omnLDbar  =  \f{\omnLD}{2 \zeta}, \qquad
\kmLbar = \f{\kmL}{2 \zeta}, \qquad 
\omLbar = \f{\omL}{2 \zeta},
\eel 
which satisfy the following normalization conditions:
\bel{rescalednoem}
\f{1}{2} \omnLbar \omnUbar = 1, \qquad
\f{1}{2} {\omnLDbar} {\omnUDbar} = -1, \qquad 
\kbar \cdot \kbar - \obar \cdot \obar = 1 \,.
\eel
%

\subsection{Observables}
\label{sec:obs}

The matrix distribution functions, given in \rftwo{fplusrsxp}{fminusrsxp}, can be used to obtain the energy momentum tensor~\cite{deGroot:1980}
\bel{Tmn0}
\TmnU &=&  \kappa \int \f{d^3p}{2 E_p}  \pmu \pnu  {\trf} \left( X^+ + X^- \right)   
\eel
and the spin tensor \cite{Becattini:2009wh}
\bel{st110}
\slmnU = \kappa \int \f{d^3p}{2 E_p} \, p^\lambda \, {\trf} \left[(X^+\!-\!X^-) \SmunuU \right] .
\eel
Here $\kappa = g/(2\pi)^3$ with $g$ accounting for internal degrees of freedom different from spin (for example, color or isospin). 

\begin{figure}[t!]
\includegraphics[angle=0,width=0.6\textwidth]{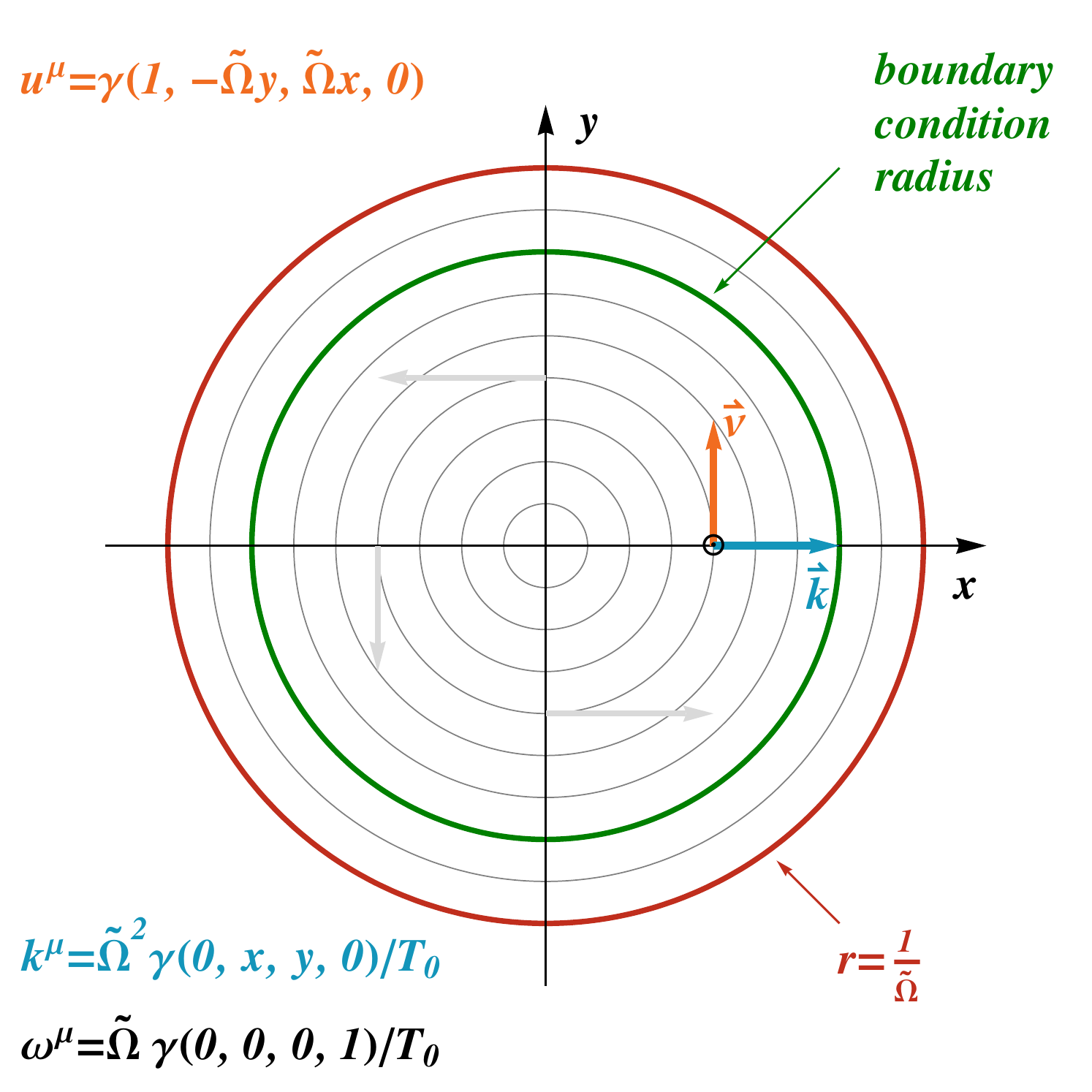} 
\caption{(Color online)  Hydrodynamic flow and polarization variables for the global thermodynamic equilibrium state studied in~\cite{Becattini:2007sr}.
}
\label{fig:vortex}
\end{figure}

For the discussion of the PL four-vector in Sec.~\ref{sec:PL} it is convenient to introduce the total particle current
\bel{calNmu0}
{\cal N}^\mu &=&  \kappa \int \f{d^3p}{2 E_p}  \pmu \left[ \trf( X^+ ) + \trf ( X^- )  \right] \, ,
\eel
which sums the contributions from particles and antiparticles, and the net conserved charge current
\bel{Nmu0}
N^\mu &=&  \kappa \int \f{d^3p}{2 E_p}  \pmu \left[ \trf( X^+ ) - \trf ( X^- )  \right] \, ,
\eel
which is the difference between the particle and antiparticle currents.

\subsection{Stationary vortex}
\label{sec:vortex}

In Fig.~\ref{fig:vortex} we show the vectors $\ev, \bv, \kv$, and $\ov$ for the stationary vortex studied in Ref.~\cite{Becattini:2007sr}. In this case $\ev=(0,0,0)$ and $\bv=(0,0,\Ot/T_0)$, where $\Ot$ and $T_0$ are constant parameters corresponding to the angular momentum and central temperature of the vortex. The hydrodynamic flow is given by the  four-vector $u^\mu$~\cite{Florkowski:2017ruc}, 
\bel{uvortex}
\hspace{-0.5cm} u^0 = \gamma, \quad u^1 = - \, \gamma \, \Ot \, y, \quad u^2 = \gamma \, \Ot \, x, \quad u^3 = 0,
\eel
where $\gamma = 1/\sqrt{1 - \Ot^2 r^2}$, and $r= \sqrt{x^2 + y^2}$  is the distance from the  center of the vortex in the transverse plane. Since the flow velocity cannot exceed the speed of light, the flow profile \rf{uvortex} may be realized only within a cylinder of radius $R < 1/\Ot$ (illustrated by the green circle in Fig.~\ref{fig:vortex}). In the present case, $\kv = \Ot^2 (\gamma/T_0) (x,y,0)$, $\ov=\gamma \bv = \Ot (\gamma/T_0) (0,0,1)$, while $\ev = 0$.

\section{Spin polarization three-vector}
\label{sec:P3V}

\subsection{Decomposition in terms of Pauli matrices}
\label{sec:Pmdec}

We expect that the spin observables are represented by Pauli matrices $\sigv$ and that the expectation values of $\sigv$ provide information on the polarization of spin-$\onehalf$ particles in their rest frame. Since we consider Dirac bispinors obtained by so called {\it canonical Lorentz boosts} applied to states with zero momentum, we refer to the resulting spin distributions and particle rest frames as the canonical ones (differing from other definitions by a rotation).

In the following, we start with \rftwo{fplusrsxp}{fminusrsxp}, and derive a decomposition of the distribution functions $f^\pm(x,p)$ in terms of Pauli matrices. The decomposition introduces a polarization vector $\Pv(x,p)$, which can be interpreted as a spatial part of the polarization four-vector $P^\mu(x,p)=(0,\Pv)$, with a vanishing zeroth component. The average polarization vector $\calPv(x,p)$ is normalized by the trace of the distribution functions. In Sec.~\ref{sec:PL} we demonstrate that $\calPv(x,p)$ agrees with the spatial part of the PL four-vector $\left(\pi^0(x,p), \piv(x,p)\right)$, obtained from the spin tensor employed in \cite{Florkowski:2017ruc}, boosted to the canonical rest frame of particles with the  LAB four-momentum $p^\mu$.

Using \rftwo{XpmM}{Mpmexp}, the spin dependent distribution functions given in \rftwo{fplusrsxp}{fminusrsxp}, can be rewritten in a form linear in the Dirac spin tensor,
\bel{fplusrsxp1}
\fplusrsxp =  e^{\xi - p \cdot \beta} \left[ \cosh(\zeta) \delta_{rs} + \f{\sinh(\zeta)}{2\zeta}  \, \ubarrp \oabL \SabU \usp \right],
\eel
\bel{fminusrsxp1}
\fminusrsxp = e^{-\xi - p \cdot \beta} \left[ \cosh(\zeta) \delta_{rs} + \f{\sinh(\zeta)}{2\zeta}  \,  \vbarsp \oabL \SabU \vrp \right].
\eel
%
To proceed further we use the two identities:
\beq 
\ubarrp   \S0iU  \usp  &=&   \vbarsp   \S0iU  \vrp = -\frac{1}{2 m} \epsLijk p^j \sigUk \, ,
\label{id1} \\
\ubarrp   \SijU  \usp   &=&   \vbarsp  \SijU  \vrp  =
\frac{E_p}{2 m} \epsLijk \lp \sigUk - p^k \frac{\pv\!\cdot\!\sigvrs}{E_p (E_p+m)}\rp. \label{id2}
\eeq
%

Using \rftwo{id1}{id2}, a straightforward calculation yields 
\bel{uoSu}
\ubarrp \oabL \SabU  \usp =  \vbarsp \oabL \SabU  \vrp  = P^0 \delta_{rs} - \Pv \cdot \sigvrs ,
\eel
where $P^0 = 0$ and the three-vector $\Pv$ is given by
\beq
\Pv &=& \f{1}{m} \left[ \vphantom{\f{1}{m} } 
u^0 \lp E_p  \ov - \pv \times \kv - \frac{ \pv \!\cdot\! \ov}{E_p + m}\,\pv \rp  
- \, \omega^0 \lp E_p  \uv  - \frac{\pv \!\cdot\! \uv}{E_p+m} \, \pv \rp  
 \right. \nn \\
&&   \left. \hspace{1cm}  +  \, k^0 (\pv \times \uv)  +   (\pv\!\cdot\! \ov)\uv - (\pv\cdot \uv)\ov 
-  \left( E_p \, (\kv\times\uv) - \frac{ \pv\!\cdot\!(\kv\times\uv)}{E_p+m} \, \pv \right)
 \right]
\label{Avkom}
\eeq
or
\beq
\Pv &=& \f{1}{m} \left[  E_p \, \bv - \pv \times \ev - \f{\pv \cdot \bv}{E_p + m} \pv \right],
\label{Aveb}
\eeq
depending whether we use the parameterisation given in \rf{omunuL} or \rf{omeb}, respectively. We note that the expression on the right-hand side of \rf{Aveb} is just the field $\bv$ in the particle rest frame~\cite{Jackson:1998}. 
We summarize this finding by writing
\beq
\Pv &=&  \bv_\ast .
\label{AvebPRF}
\eeq
Thus, the polarization is determined by the field $\bv$ in the canonical particle rest frame of the particle. 

Using \rf{uoSu} in \rftwo{fplusrsxp1}{fminusrsxp1} we then find
\bel{fpm}
f^\pm(x,p) =  e^{\pm \xi - p \cdot \beta} \left[ \cosh(\zeta)  - \f{\sinh(\zeta)}{2\zeta}  \, \Pv \cdot \sigv \right].
\eel
In the next step, we define the average polarization vector $\calPv$ by the formula
\beq
\calPv =  \f{1}{2} \f{ \trt \left[ (f^+ + f^-) \sv\right]  }{\trt \left[ f^+ + f^- \right] }  = -\f{1}{2} \tanh(\zeta) \Pbarv 
\eeq
where we have introduced the notation
\bel{Pbar}
\Pbarv = \f{\Pv}{2\zeta}. 
\eel
Using \rf{zeta}, we obtain an alternative expressions
\beq
\calPv = - \f{1}{2} \tanh\left[  \f{1}{2} \sqrt{ \bv_\ast \cdot \bv_\ast - \ev_\ast \cdot \ev_\ast } \right]  
\f{ \bv_\ast }{\sqrt{ \bv_\ast \cdot \bv_\ast - \ev_\ast \cdot \ev_\ast }}  ,
\eeq
where we have used the property that the quantity $\bv \cdot \bv - \ev \cdot \ev$ is independent of the choice of the Lorentz frame.

\section{Pauli-Luba\'nski four-vector}
\label{sec:PL}

\subsection{Phase-space density}
\label{sec:PLden}

Starting from the definition of the Pauli-Luba\'nski four-vector $\Pi_\mu$, and following the method introduced in Ref.~\cite{Becattini:2013fla}, we introduce the phase-space density of $\Pi_\mu$ defined by the following expression
\bel{PL1}
E_p \f{d \Delta \Pi_\mu(x,p)}{d^3p}  = -\f{1}{2} \epsLmnab \, \Delta \Sigma_\lambda(x) \, 
E_p \f{d J^{\lambda, \nu\alpha}(x,p)}{d^3p}
\f{p^\beta}{m}.
\eel
Here $\Delta \Sigma_\lambda$ denotes a space-time element of the fluid and $E_p d J^{\lambda, \nu\alpha}/d^3p$ denotes the invariant phase-space density of the angular momentum of particles with four-momentum $p$. Using definitions introduced in Ref.~\cite{Florkowski:2017ruc}, analogous to \rftwo{Tmn0}{st110}, we find
\beq
E_p \f{d J^{\lambda, \nu\alpha}(x,p)}{d^3p} &=&   \f{\kappa}{2} \, p^\lambda \left(x^\nu p^\alpha  - x^\alpha p^\nu \right)
 \trf (X^+ + X^-)  + \f{\kappa}{2} \, p^\lambda \trf \left[ \left(X^+ - X^-\right) \Sigma^{\nu \alpha} \right].
\label{Jlna1}
\eeq
Clearly, the orbital part in \rf{Jlna1} does not contribute to the density of $\Pi_\mu$. Hence we find
\bel{PL2}
E_p \f{d \Delta \Pi_\mu(x,p)}{d^3p}  = -\f{1}{2} \epsLmnab \, \Delta \Sigma_\lambda(x) \, 
E_p \f{d S^{\lambda, \nu\alpha}(x,p)}{d^3p}
\f{p^\beta}{m},
\eel
where
\beq
E_p \f{d S^{\lambda, \nu\alpha}(x,p)}{d^3p} &=&  
\f{\kappa}{2} \, p^\lambda \trf \left[ \left(X^+ - X^-\right) \Sigma^{\nu \alpha} \right].
\label{Slna1}
\eeq
Performing the traces in \rf{Slna1}, one obtains
\beq
E_p \f{d S^{\lambda, \nu\alpha}(x,p)}{d^3p} &=&  \, \kappa \, e^{- p \cdot \beta} \cosh(\xi)
\f{ \sinh(\zeta)}{ \zeta} p^\lambda \omega^{\nu \alpha}.
\label{Slna2}
\eeq
Now, substituting \rf{Slna2} into \rf{PL2} we find
\beq
E_p \f{d \Delta \Pi_\mu(x,p)}{d^3p}  &=& -\f{1}{2} \epsLmnab \, \Delta \Sigma_\lambda(x) \, 
\f{w(x,p)}{4 \zeta} p^\lambda \omega^{\nu \alpha} \f{p^\beta}{m} \nn \\
&=& - \, \Delta \Sigma \cdot  p \, 
\f{w(x,p)}{4 m \zeta}  \, {\tilde \omega}_{\mu \beta} \, p^\beta.
\label{PL3} 
\eeq
where $w(x,p) = 4 \kappa \, e^{- p \cdot \beta} \cosh(\xi) \sinh(\zeta)$. Since we are interested in the polarization effect per particle, it is convenient to introduce the particle density in the volume $\Delta \Sigma$ defined with the help of \rf{calNmu0}. This leads to the expression
\beq
E_p \f{d \Delta {\cal N}}{d^3p} &=& \f{\kappa}{2} \, \Delta \Sigma \cdot  p \, \trf \, \left(X^+ + X^-\right) 
= 4 \kappa  \, \Delta \Sigma \cdot  p \, e^{-p \cdot \beta} \cosh(\xi) \cosh(\zeta).
\label{DcalN}
\eeq
The PL vector per particle is then obtained by dividing \rf{PL3} by \rf{DcalN},
\bel{PL4}
\pi_\mu(x,p) \equiv \f{\Delta \Pi_\mu(x,p)}{\Delta {\cal N}(x,p)} &=& 
-\f{\tanh(\zeta)}{4 m \zeta} \, {\tilde \omega}_{\mu \beta} \, p^\beta \nn \\
&=& -\f{\tanh(\zeta)}{4 m \zeta}
\left( \omega_\mu \, p \cdot u - u_\mu \, p \cdot \omega +
\epsilon_{\mu \rho \sigma \beta} k^\rho u^\sigma p^\beta \right),
\eel
where in the second line we have used the definition of the dual polarization tensor given in \rf{omunuLD}. Using the rescaled quantities, defined in \rf{rescaled}, we finally arrive at
\bel{PL5}
\pi_\mu(x,p) &=& 
-\f{\tanh(\zeta)}{2 m} \, {\bar {\tilde \omega}}_{\mu \beta} \, p^\beta = -  \f{\tanh(\zeta)}{2 m}
\left( {\bar \omega}_\mu \, p \cdot u - u_\mu \, p \cdot {\bar \omega} +
\epsilon_{\mu \rho \sigma \beta} {\bar k}^\rho u^\sigma p^\beta \right).
\eel

\subsection{Boost to particle rest frame}
\label{sec:pboost}

In order to boost the four-vector $\pi^\mu$ to the local rest frame of a particle with momentum $p$, we use the Lorentz transformation for a canonical boost~\cite{Leader:2001}
%
\bel{boost}
\Lambda^\mu_{\,\,\,\nu}(-\vv_p) = 
\begin{bmatrix}
\f{E_p}{m} & -\f{p^1}{m} & -\f{p^2}{m} & -\f{p^3}{m} \\
-\f{p^1}{m} & \,\,\,1 + \alpha_p p^1 p^1 &  \alpha_p p^1 p^2 &  \alpha_p p^1 p^3 \\
-\f{p^2}{m} &  \alpha_p p^2 p^1 & \,\,\, 1+ \alpha_p p^2 p^2 &  \alpha_p p^2 p^3 \\
-\f{p^3}{m} & \alpha_p p^3 p^1 &  \alpha_p p^3 p^2 & \,\,\, 1+ \alpha_p p^3 p^3 \\
\end{bmatrix},
\eel
where $\vv_p = \pv/E_p$ and $\alpha_p = 1/(m (E_p + m))$. Using \rf{PL5}, we can express the time and space components of $\pi^\mu = (\pi^0, \piv)$ in the LAB frame in the three-vector notation
\bel{pi0}
\pi^0 =  -  \f{\tanh\zeta}{4 \zeta m} 
\left(u^0 \pv \cdot \ov - \omega^0 \pv \cdot \uv  + \kv \cdot (\pv \times \uv)  \right),
\eel
\bel{piv}
\piv = -  \f{\tanh\zeta}{4 \zeta m} \left( \ov \, p \cdot u - \uv \, p \cdot \omega + k^0 \, \pv \times \uv
- u^0 \pv \times \kv - E_p \, \kv \times \uv  \right).
\eel
By applying the Lorentz transformation \rf{boost} to \rf{pi0} and \rf{piv} we find
\bel{pi0PRF}
\pi^0_\ast   = 0
\eel
and
\bel{pivPRF}
\piv_\ast   = \calPv =  - \f{1}{2} \tanh(\zeta) \Pbarv \, .
\eel
Due to the Lorentz four-vector character of $\pi_\mu$, we have $\pi_\mu \pi^\mu = \pi_\mu^\ast \pi^\mu_\ast= - \calPv^2$.

\section{Other spin-tensor forms}
\label{sec:Slmn}

\subsection{Independence of the PL four-vector}
\label{sec:SlmnPL}

Another form for the spin tensor, which can be used to construct the PL four-vector, is given by
\beq
\slmnU_{\rm can} &=&  \kappa \int \f{d^3p}{2 E_p} 
\left( p^\lambda \trf \left[(X^+-X^-) \Sigma^{\mu\nu}  \right] 
- \pmu  \trf \left[(X^+-X^-) \Sigma^{\lambda \nu}\right]  \right. \nn \\
&& \left. \hspace{3cm} + \pnu  \trf \left[(X^+-X^-) \Sigma^{\lambda \mu}\right]  \right).
\label{SlmnCan}
\eeq
Equation \rfn{SlmnCan} was derived in Ref.~\cite{Becattini:2013fla} and corresponds to the canonical spin tensor, obtained directly by applying Noether's theorem to the Dirac lagrangian. This form of the spin tensor differs from \rf{st110} by two additional terms containing $p^\mu$ and $p^\nu$ in the integrand. We note that the two additional terms in the integrand do not contribute to the PL four-vector, since they vanish if contracted with the Levi-Civita tensor in \rf{PL2}.

Yet another version of the spin tensor, introduced in the textbook by Groot, Leeuven, and Weert \cite{deGroot:1980}, reads
\beq
\slmnU_{\rm GLW} &=&  \kappa  \int \f{d^3p}{E_p} p^\lambda
\left(  \trt\left[ f^+(x,p) \Sigma^{\mu\nu}_+ \right] + \trt\left[ f^-(x,p) \Sigma^{\mu\nu}_- \right] \right),
\label{SlmnGLW1}
\eeq
where
\bel{Spm}
\left[\Sigma^{\mu\nu}_+ \right]_{rs} = \ubarrp \SmunuU \usp, \quad
\left[\Sigma^{\mu\nu}_- \right]_{rs} = \vbarsp \SmunuU \vrp.
\eel
By changing the trace over spin indices to the spinor trace and using the commutation relation
\bel{com1}
\left[ \SmnU,  p_\alpha \gamma^\alpha \right] = i p^\nu \gamma^\mu - i p^\mu \gamma^\nu
\eel
we find
\beq
\slmnU_{\rm GLW} &=&  \kappa \int \f{d^3p}{2 E_p} p^\lambda
\left(   \trf \left[ (X^+-X^-) \Sigma^{\mu\nu} \right]  \vphantom{ \f{i}{2 m^2}} \right. \nn \\
&& \left. \hspace{1cm}
+ \f{i}{2 m^2} \trf \left[ (X^+-X^-) p_\alpha \gamma^\alpha \left( \gamma^\mu p^\nu - \gamma^\nu p^\mu \right) \right] 
 \right).
\label{SlmnGLW2}
\eeq
We again notice that only the first term of the integrand in \rf{SlmnGLW2} contributes to the PL four-vector. Interestingly, the resulting PL four vector is identical for all three forms of the spin tensor.

\subsection{Large $m/T$ limit of the Groot-Leeuven-Weert  spin tensor}
\label{sec:SlmnGLW}

In this section we consider the large $m/T$ limit of the spin tensor introduced in Ref.~\cite{deGroot:1980}. This exercise is instructive, since the result is simple and sheds some light on the relevance of the Frenkel condition. We introduce the symbol $S^{\lambda, \mu \nu}_\Delta$ for the second term in \rf{SlmnGLW2}:
\bel{SlmnGLW3}
\slmnU_{\rm GLW} &=& \slmnU + S^{\lambda, \mu \nu}_\Delta.
\eel
Then, using the identity
\bel{trace}
\trf \left[(a + b \, \omega_{\rho \sigma} \Sigma^{\rho \sigma}) \, 
p_\alpha   \gamma^\alpha \, (\gamma^\mu p^\nu - \gamma^\nu p^\mu ) \right]
= 4 i \,b\,p_\alpha \left( p^\nu \omega^{\mu \alpha} - p^\mu \omega^{\nu \alpha} \right)
\eel
in \rf{SlmnGLW2}, with $a$ and $b$ being arbitrary scalars, we find
\begin{eqnarray}
\label{sdelta}
S^{\lambda, \mu \nu}_\Delta
&=&  \f{\kappa  \sinh(\zeta) \cosh(\xi)}{m^2 \, \zeta}  \,  \int \f{d^3p}{E_p}  \, e^{- \beta \cdot p} \,
p^\lambda \, \, p_\alpha (p^\mu \omega^{\nu\alpha} -p^\nu \omega^{\mu\alpha}) .
\end{eqnarray}
The integral over momentum in \rf{sdelta} can be tensor decomposed in a combination of terms containing the four-velocity $\umU$ and the metric tensor $g^{\mu\nu}$. After contraction with the polarization tensor $\omnU$, this leads to the expression
\begin{eqnarray}
\label{sdeltaf}
S^{\lambda, \mu \nu}_\Delta
&=&  \f{g \sinh(\zeta) \cosh(\xi)}{m^2 \, \zeta}  \, T\, \left[ \e0(T) + \P0(T) \right] s^{\lambda, \mu \nu}_\Delta \nn \\
&& + \, 
\f{g \sinh(\zeta) \cosh(\xi)}{T \, \zeta}   \, \P0(T) \left( k^\nu u^\lambda u^\mu - k^\mu u^\lambda u^\nu \right).
\end{eqnarray}
Here $\e0(T)$ and $\P0(T)$ are, respectively, the energy density and pressure of classical spinless particles with the mass $m$ computed at the temperature $T$~\cite{Florkowski:2017ruc}~\footnote{Thermodynamic functions, such as $\e0(T)$ or $\P0(T)$, include the factor $(2\pi)^3$ in the denominator of the momentum integration measure, hence the factor $\kappa$ in \rf{sdeltaf} has been replaced by $g$.}, and the tensor $s^{\lambda, \mu \nu}_\Delta$ is defined by
\bel{br}
s^{\lambda, \mu \nu}_\Delta = 2 u^\lambda\,( \omega^{\mu\nu}  + 3\,( k^\nu  u^\mu - \,k^\mu u^\nu ))  + \omega^{\mu\lambda} u^\nu - \omega^{\nu\lambda} u^\mu +  k^\mu g^{\lambda\nu} - k^\nu g^{\lambda\mu} .
\eel

For classical statistics used in this work $\P0(T) = T \,\n0(T)$. Moreover, in the limit $m \ll T$, we have $\e0(T) \approx m \,\n0(T)$. Thus, the first term on the right-hand side of  \rf{sdeltaf} is of order $T/m$  compared to the second one, and thus negligible in the large $m/T$ limit. Moreover, the term in the second line of  \rf{sdeltaf} cancels exactly the part of $\slmnU$ depending on the four-vector $k$, see Eqs.~(5) and (27) in \cite{Florkowski:2017ruc}. Consequently, the large $m/T$ limit of the definition in \rf{SlmnGLW1} has the form
\bel{SlmnGLW4}
\slmnU_{\rm GLW} &=& g \, \f{\sinh(\zeta) } {\zeta} \, \cosh(\xi) \, \n0(T) \, u^\lambda \,\epsilon^{\mu\nu \alpha \beta} u_\alpha \omega_\beta
\eel
which is independent of $k$. Interestingly, this result is similar to imposing the Frenkel condition $k^\mu=0$.

\section{Summary and conclusions}
\label{sec:SC}

In this work we have studied properties of the spin density matrices used in recent formulations of relativistic hydrodynamics of particles with spin $\onehalf$. We showed that the total polarization vector, obtained by calculating the trace of the product of spin density matrices and the Pauli matrices, agrees with the Pauli Luba\'nski four-vector obtained from the spin tensor used in Ref.~\cite{Florkowski:2017ruc}. This allows for a natural determination of the spin polarization vectors of particles and antiparticles. We have also demonstrated, that the two other forms of the spin tensor yield the same polarization vector validating that the form used in \cite{Florkowski:2017ruc} represents an appropriate classical approximation for the spin tensor. 

\begin{acknowledgments}
We thank Leonardo Tinti and Giorgio Torrieri for clarifying discussions. This work was supported in part by the DFG through the grant CRC-TR 211. W.F. and R.R. were supported in part by the Polish National Science Center Grant No. 2016/23/B/ST2/00717 and by the ExtreMe Matter Institute EMMI at the GSI Helmholtzzentrum f\"ur Schwerionenforschung, Darmstadt, Germany. A.J. was supported in part by the DST-INSPIRE faculty research grant and by the ExtreMe Matter Institute EMMI at GSI. E.S. was supported by BMBF Verbundprojekt 05P2015 - Alice at High Rate. 
\end{acknowledgments}

\appendix
%

\section{Dirac spinors}
\label{sec:DS}

The conventions for labels and signs of bispinors used in this work are as follows:
\beq
\usp =  \sqrt{\frac{E_p+m}{2m}} 
\lp\begin{array}{cc}   1 & \varphi_s\\ \frac{\sv \cdot \pv}{E_p+m} &  \varphi_s   \end{array}\rp, \qquad 
\vsp =  \sqrt{\frac{E_p+m}{2m}} 
\lp\begin{array}{cc}   \frac{\sv \cdot \pv}{E_p+m} & \chi_s\\ 1 &  \chi_s   \end{array}\rp,
\eeq
with 
\beq
\varphi_1 =    
\lp\begin{array}{c}   1  \\ 0 \end{array}\rp,  \qquad
\varphi_2 =    
\lp\begin{array}{c}   0  \\ 1 \end{array}\rp, \qquad
\chi_1 =    
\lp\begin{array}{c}   0  \\ 1 \end{array}\rp, \qquad
\chi_2 =    
-\lp\begin{array}{c}   1  \\ 0 \end{array}\rp.
\eeq
The spin operator $\SmunuU$ is defined by the expression
\bel{SDO}
\SmunuU  = \f{1}{2} \sigma^{\mu\nu} = \f{i}{4} [\gamma^\mu,\gamma^\nu], 
\eel
which in the Dirac representation gives
\beq
\Sigma^{0i} = \f{i}{2} \lp\begin{array}{cc} 
0 & \sigma^i \\ \sigma^i & 0 
\end{array}\rp, \qquad 
\Sigma^{ij} = \f{1}{2} \epsilon_{ijk} \lp\begin{array}{cc} \sigma^k & 0\\ 0 & \,\,\,\sigma^k \end{array}\rp,
\eeq
with $\sigma^i$ being the $i$th Pauli matrix. 

\section{Spin matrices $M^{\pm}$}
\label{sec:spinden}
%
In this section we present details of the calculation of the matrix $M^{\pm}$, which is defined by the exponential function of the Dirac spin operator, see \rf{Mpm}. To do this calculation most easily we first switch to the chiral representation of the Dirac matrices, where $\SmunuU$ is block diagonal, and then move to the local rest frame of the fluid element, where $u^{\mu}_\circ~=~\Lambda^{\mu}_{\,\,\,\nu}~u^{\nu}~=~(1,0,0,0)$. Calculation of the exponential function in \rf{Mpm} in the chiral representation with $u^{\mu}_\circ~=~(1,0,0,0)$ is reduced to the well known calculation of the exponential function of a linear combination of the Pauli matrices. Once it is done, we come back to the LAB frame (from the local rest frame of the fluid element) and perform a unitary transformation back to the Dirac representation.

With $\SL$ denoting the transformation matrix that corresponds to the Lorentz transformation $\Lambda$, we have
\beq
& \SL \gmU \SL^{-1} = \Lambda^{\mu}_{\,\,\,\nu}\gamma^{\nu},
\eeq
\beq
& \SL \SmnU \SL^{-1} = \Lambda^{\mu}_{\,\,\,\alpha}\Lambda^{\nu}_{\,\,\,\beta} \Sigma^{\alpha \beta},
\eeq
and
\beq
M^{\pm}_\circ &=& \SL M^\pm\SL^{-1} = \SL \exp\lp \pm \frac{1}{2} \omnL \SmnU  \rp\SL^{-1} \nn \\
&=&  \exp\lp \pm \frac{1}{2} \omnL \SL\SmnU  \SL^{-1}\rp  
=\exp\lp \pm \frac{1}{2} \omega^\circ_{\mu \nu}   \Sigma^{\mu \nu} \rp.
\eeq
Working in the chiral representation, we use
\beq
\Sigma^{0i} = \f{i}{2} \lp\begin{array}{cc} \sigma^i & 0\\ 0 & -\sigma^i \end{array}\rp, \qquad 
\Sigma^{ij} = \f{1}{2} \epsilon_{ijk} \lp\begin{array}{cc} \sigma^k & 0\\ 0 & \,\,\,\sigma^k \end{array}\rp.
\eeq
In the fluid rest frame $\omega^\circ_{0i}=k_\circ^i$ and $\omega^\circ_{ij}=-\epsilon_{ijk} \omega_\circ^k$\,, thus we have
\beq
& &  \pm \frac{1}{2} \omega^\circ_{\mu \nu}   \Sigma^{\mu \nu}   
=  \lp\begin{array}{cc}   \pm\zv\cdot\sv & 0\\ 0 &  \pm(\zv\cdot\sv)^\dagger  \end{array}\rp, 
\label{Mapp1} 
\eeq
where ${\zv}=(-\ov_\circ+ i\kv_\circ)/2$. Consequently, using the method for exponentiating the Pauli matrices we obtain
\beq
M^{\pm}_\circ &=& \exp\lsb  \lp\begin{array}{cc}   \pm\zv\cdot\sv & 0\\ 0 &  \pm(\zv\cdot\sv)^\dagger  \end{array}\rp\rsb \nn \\
&=&     \lp\begin{array}{cc}  \cosh z \pm \frac{\sinh z}{z} \, \zv\cdot\sv & 0\\ 0 &  \cosh z^\star \pm \frac{\sinh z^\star}{z^\star}\,(\zv\cdot\sv)^\dagger  \end{array}\rp,
\label{Mapp2} 
\eeq
with   $z^2 = \zv \cdot\zv$. Introducing the $\gamma_5$  matrix in the chiral representation and using Eq.~(\ref{Mapp1}) one can further simplify Eq.~(\ref{Mapp2})  to
\beq
& &M^{\pm}_\circ =     \mathbb{1} \lp \Re(\cosh z) \pm \Re\lp\frac{\sinh z}{z}\rp\frac{1}{2}\omega_{\mu \nu}^\circ  \Sigma^{\mu \nu}\rp\nonumber \\
&&\qquad\qquad+ i \gamma_5 \lp \Im(\cosh z)\pm \Im\lp\frac{\sinh z}{z}\rp\frac{1}{2}\omega^\circ_{\mu \nu}   \Sigma^{\mu \nu}\rp.
\label{Mapp3} 
\eeq
As this equation is manifestly Lorentz covariant, we may drop the symbol $\circ$ denoting that it has been derived in the local fluid rest frame. Moreover, as it has a form expressed in terms of the Dirac matrices, it is valid in any representation, including the Dirac one. 


\bibliography{hydro_review}{}
\bibliographystyle{utphys}

\newpage

\end{document}